\documentclass[prf,aps,psfig,showpacs,nofootinbib,nobibnotes,superscriptaddress,preprintnumbers,times]{revtex4}
\usepackage{graphicx}
\usepackage{bm,color}

\newcommand{\bra}[1]{\langle #1\rangle}
\newcommand{\nab}{\mbox{\boldmath $\nabla$} {}}
\newcommand{\ii}{{\rm{i}}}
\newcommand{\dd}{{\rm{d}}}

\newcommand{\Lu}{{\rm{Lu}}}
\newcommand{\Rey}{{\rm{Re}}}

\newcommand{\be}{\begin{equation}}
\newcommand{\ee}{\end{equation}}

\newcommand{\bea}{\begin{eqnarray}}
\newcommand{\eea}{\end{eqnarray}}
\newcommand{\bean}{\begin{eqnarray*}}
\newcommand{\eean}{\end{eqnarray*}}

\def\xiM{\xi_{\rm M}}
\def\xiK{\xi_{\rm K}}
\def\sigmaM{\sigma_{\rm M}}
\def\sigmaK{\sigma_{\rm K}}

\newcommand{\SSSS}{\mbox{\boldmath ${\sf S}$} {}}
\newcommand{\AAA}{\bm{A}}
\newcommand{\ggg}{\bm{g}}
\newcommand{\BB}{\bm{B}}
\newcommand{\xx}{\bm{x}}
\newcommand{\meanBB}{\overline{\bm{B}}}
\newcommand{\meanJJ}{\overline{\bm{J}}}
\newcommand{\JJ}{\bm{J}}
\newcommand{\PP}{\bm{P}}
\newcommand{\uu}{\bm{u}}
\newcommand{\kk}{\bm{k}}
\newcommand{\oo}{\bm{\omega}}
\def\cs{c_{\rm s}}

\def\km{k_{\rm m}}
\def\Pm{{\rm Pr}_{\rm M}}
\def\Rm{{\rm Re}_{\rm M}}

\def\urms{u_{\rm rms}}

\def\Brms{B_{\rm rms}}
\def\etat{\eta_{\rm t}}
\def\etaT{\eta_{\rm T}}
\def\epsf{\epsilon_{\rm f}}
\def\EEK{{\cal E}_{\rm K}}
\def\EEM{{\cal E}_{\rm M}}

\def\EK{E_{\rm K}}
\def\EM{E_{\rm M}}
\def\HK{H_{\rm K}}
\def\HM{H_{\rm M}}
\def\la{\mathrel{\mathchoice {\vcenter{\offinterlineskip\halign{\hfil
$\displaystyle##$\hfil\cr<\cr\sim\cr}}}
{\vcenter{\offinterlineskip\halign{\hfil$\textstyle##$\hfil\cr<\cr\sim\cr}}}
{\vcenter{\offinterlineskip\halign{\hfil$\scriptstyle##$\hfil\cr<\cr\sim\cr}}}
{\vcenter{\offinterlineskip\halign{\hfil$\scriptscriptstyle##$\hfil\cr<\cr\sim\cr}}}}}
\def\ga{\mathrel{\mathchoice {\vcenter{\offinterlineskip\halign{\hfil
$\displaystyle##$\hfil\cr>\cr\sim\cr}}}
{\vcenter{\offinterlineskip\halign{\hfil$\textstyle##$\hfil\cr>\cr\sim\cr}}}
{\vcenter{\offinterlineskip\halign{\hfil$\scriptstyle##$\hfil\cr>\cr\sim\cr}}}
{\vcenter{\offinterlineskip\halign{\hfil$\scriptscriptstyle##$\hfil\cr>\cr\sim\cr}}}}}

\newcommand{\Sec}[1]{Sect.~\ref{#1}}

\newcommand{\Fig}[1]{Fig.~\ref{#1}}

\newcommand{\Figp}[2]{Fig.~\ref{#1}({#2})}
\newcommand{\Figsp}[3]{Figs.~\ref{#1}({#2}) and ({#3})}

\newcommand{\Figs}[2]{Figs.~\ref{#1} and \ref{#2}}

\newcommand{\Tab}[1]{Table~\ref{#1}}

\newcommand{\ybook}[3]{ {\em #2}.\ #3 (#1).}

\newcommand{\arXiv}[3]{, #2, arXiv:#3 (#1).}

\newcommand{\ymn}[5]{, #5, {\rm Month. Not. Roy.\ Astron.\ Soc.\ }{\bf #2}, #3--#4 (#1).}

\newcommand{\ymhd}[5]{, #5, {\rm Magnetohydrohydrodyn. }{\bf #2}, #3--#4 (#1).}

\newcommand{\ynatcN}[4]{, #4, {\rm Nat. Comm. }{\bf #2}, #3 (#1).}

\newcommand{\yjfm}[5]{, #5, {\rm J.\ Fluid Mech.\ }{\bf #2}, #3--#4 (#1).}

\newcommand{\yprd}[5]{, #5, {\rm Phys.\ Rev.\ D }{\bf #2}, #3--#4 (#1).}

\newcommand{\yprdN}[4]{, #4, {\rm Phys.\ Rev.\ D }{\bf #2}, #3 (#1).}

\newcommand{\ypreN}[4]{, #4, {\rm Phys.\ Rev.\ E }{\bf #2}, #3 (#1).}

\newcommand{\yprl}[5]{, #5, {\rm Phys.\ Rev.\ Lett.\ }{\bf #2}, #3--#4 (#1).}
\newcommand{\yprlN}[4]{, #4, {\rm Phys.\ Rev.\ Lett.\ }{\bf #2}, #3 (#1).}

\newcommand{\yptrsa}[5]{, #5, {\rm Phil.\ Trans.\ Roy.\ Soc.\ A }{\bf #2}, #3--#4 (#1).}

\newcommand{\yapj}[5]{, #5, {\rm Astrophys.\ J.\ }{\bf #2}, #3--#4 (#1).}
\newcommand{\yapjN}[4]{, #4, {\rm Astrophys.\ J.\ }{\bf #2}, #3 (#1).}

\newcommand{\yrppN}[4]{, #4, {\rm Rep.\ Prog.\ Phys.\ }{\bf #2}, #3 (#1).}

\newcommand{\yppN}[4]{, #4, {\rm Phys.\ Plasmas }{\bf #2}, #3 (#1).}
\newcommand{\yppcf}[5]{, #5, {\rm Plasmas Phys. Contr. Fusion }{\bf #2}, #3--#4 (#1).}

\newcommand{\ypf}[5]{, #5, {\rm Phys.\ Fluids }{\bf #2}, #3--#4 (#1).}

\newcommand{\yjour}[6]{, #6, {\rm #2} {\bf #3}, #4--#5 (#1).}
\newcommand{\yjourN}[5]{, #5, {\rm #2} {\bf #3}, #4 (#1).}

\newcommand{\EQ}{\begin{equation}}
\newcommand{\EN}{\end{equation}}
\newcommand{\ba}{\begin{eqnarray}}
\newcommand{\ea}{\end{eqnarray}}

\newcommand{\Eq}[1]{Eq.~(\ref{#1})}

\def\pM{p_{\rm M}}
\def\qM{q_{\rm M}}
\def\betaM{\beta_{\rm M}}
\def\betaMz{\beta_{\rm M0}}
\def\betaMo{\beta_{\rm M1}}

\def\ii{{\rm i}}
\def\const{{\rm const}}

\def\vAz{v_{\rm A0}}

\graphicspath{{./fig/}{./png/}}

\begin{document}
\title{
The dynamo effect in decaying helical turbulence
}

\preprint{NORDITA-2017-99}

\author{Axel~Brandenburg}
\email{brandenb@nordita.org}
\affiliation{Laboratory for Atmospheric and Space Physics, University of Colorado, Boulder, CO 80303, USA}
\affiliation{JILA and Department of Astrophysical and Planetary Sciences, University of Colorado, Boulder, CO 80303, USA}
\affiliation{Nordita, KTH Royal Institute of Technology and Stockholm University,
Roslagstullsbacken 23, 10691 Stockholm, Sweden}
\affiliation{Department of Astronomy, AlbaNova University Center,
Stockholm University, 10691 Stockholm, Sweden}
\affiliation{McWilliams Center for
Cosmology and Department of Physics, Carnegie Mellon University,
5000 Forbes Ave, Pittsburgh, PA 15213, USA}

\author{Tina~Kahniashvili}
%\email{tinatin@andrew.cmu.edu}
\affiliation{McWilliams Center for
Cosmology and Department of Physics, Carnegie Mellon University,
5000 Forbes Ave, Pittsburgh, PA 15213, USA}
\affiliation{Department of Physics, Laurentian University, Ramsey
Lake Road, Sudbury, ON P3E 2C,Canada}
\affiliation{Abastumani Astrophysical Observatory, Ilia State University,
3-5 Cholokashvili St., 0194 Tbilisi, Georgia}

\author{Sayan~Mandal}
%\email{sayanm@andrew.cmu.edu}
\affiliation{McWilliams Center for
Cosmology and Department of Physics, Carnegie Mellon University,
5000 Forbes Ave, Pittsburgh, PA 15213, USA}

\author{Alberto~Roper~Pol}
%\email{Alberto.RoperPol@colorado.edu}
\affiliation{Laboratory for Atmospheric and Space Physics, University of Colorado, Boulder, CO 80303, USA}
\affiliation{Department of Aerospace Engineering Sciences, University of Colorado, Boulder, CO 80303, USA}

\author{Alexander~G.~Tevzadze}
%\email{aleko@tevza.org}
\affiliation{Faculty of Exact and Natural Sciences, Javakhishvili Tbilisi
State University, 3 Chavchavadze Ave., Tbilisi, 0179, Georgia}
\affiliation{Abastumani Astrophysical Observatory, Ilia State University, 3-5
Cholokashvili St., 0194 Tbilisi, Georgia}

\author{Tanmay~Vachaspati}
%\email{tvachasp@asu.edu}
\affiliation{Physics Department, Arizona State University,
Tempe, AZ 85287, USA}

\date{\today, $ $Revision: 1.102 $ $}

\begin{abstract}
We show that in
decaying hydromagnetic turbulence with initial kinetic helicity,
a weak magnetic field eventually becomes fully helical.
The sign of magnetic helicity is opposite to that of the kinetic
helicity---regardless of whether or not the initial magnetic field
was helical.
The magnetic field undergoes inverse cascading
with the magnetic energy decaying approximately like
$t^{-1/2}$. This is even slower than in the fully helical case,
where it decays like $t^{-2/3}$.
In this parameter range, the product of magnetic energy and correlation
length raised to a certain power slightly larger than unity, is approximately constant.
This scaling of magnetic energy persists over long time scales.
At very late times and for domain sizes large enough to accommodate
the growing spatial scales, we expect a cross-over to the $t^{-2/3}$
decay law that is commonly observed for fully helical magnetic fields.
Regardless of the presence or absence of initial kinetic helicity,
the magnetic field experiences exponential growth during the first few
turnover times, which is suggestive of small-scale dynamo action.
Our results have applications to a wide range of experimental dynamos
and astrophysical time-dependent plasmas,
including primordial turbulence in the early universe.
\end{abstract}

\pacs{47.27.-i, 47.27.nb, 47.65.Md}

\maketitle

\section{Introduction}

In electrically conducting fluids such as plasmas and liquid metals,
steady helical turbulence is known to lead to an efficient conversion of
kinetic energy into magnetic energy---a process referred to as a dynamo.
Dynamos with swirling (helical) motions can be excited at relatively small
magnetic Reynolds numbers, i.e., at moderate turbulent velocities and length
scales, as well as moderate electric conductivities \cite{Mof78,KR80}.
This is why many dynamo experiments have employed helical flows both
in the constrained and basically nonturbulent flows of the experiments performed
in Riga \cite{Riga1,Riga2} and Karlsruhe \cite{Karlsruhe1,Karlsruhe2},
as well as the unconstrained (turbulent) von K\'arm\'an flows in the experiments in
Cadarache \cite{Cadarache1,Cadarache2}.
Many other experiments are currently being worked upon
\cite{Maryland1,Maryland2,Madison1,Madison2}.
Their success is limited by the power that can be delivered by the
propellers or pumps.
A more economic type of dynamo experiment is driven by the flow that results
inside a spinning torus of liquid sodium after abruptly breaking it.
This leads to turbulence from the screw-like diverters inside the torus
\cite{Perm1,Perm2,FM17}.
Theoretical studies of laminar screw dynamos have been performed \cite{DFS03},
but the evolution of hydromagnetic turbulence is usually parameterized
in ways that ignore the effects of kinetic and magnetic helicity.

The problem of magnetic field evolution in decaying helical
turbulence in conducting media is far more general.
Neutron stars, for example, have convective turbulence during the first
minute after their formation \cite{DT92,TD93}.
The early universe could be another example of turbulence driven by expanding
bubbles after a first-order phase transition \cite{KKS86,TWW92}.
Turbulence can also be driven by magnetic fields generated at earlier
times during inflation \cite{Turner:1987bw,Ratra:1991bn}.
Transient turbulence is also being generated as a consequence of merging
galaxy clusters \cite{RBS99,RSB99}.
Even accretion discs may provide an example of decaying turbulence when
the magnetorotational instability is not excited during certain phases
\cite{OKMK08}.
A related example is that of tidal disruption events, where a star has
a close encounter with a supermassive black hole and gets disrupted.
During this process, tremendous shearing motion is being dissipated.
A fraction of it can be dissipated magnetically via strongly
time-dependent dynamo action and Joules dissipation \cite{GMcC17}.
Dynamo effects are also suspected to occur over durations
of microseconds in inertial fusion confinement plasmas
\cite{Li16,Tzeferacos17,Tzeferacos17b}.
In all these cases, one deals with decaying turbulence.
This is what makes the interpretation in terms of a dynamo effect
complicated.
Here we focus on general aspects of the dynamo mechanism rather than trying
to model specific laboratory or astrophysics conditions.

In this paper, we demonstrate for the first time that in decaying helical
turbulence, an initially nonhelical seed magnetic field undergoes a
quasi-exponential increase.
In the presence of initial kinetic helicity, this growth is followed by
a long ($\sim50,000$ turnover times) transient decay where
the magnetic energy decays like $t^{-1/2}$.
This is slower than in the case of an initially fully helical
magnetic field, which decays like $t^{-2/3}$.
It develops inverse cascade-type behavior already well before the magnetic
field becomes fully helical.
To what extent the transient decay owing to initial kinetic
helicity can be modeled in terms of advanced mean-field dynamo
theory remains open, although potentially suitable tools such as two-
and three-scale dynamo theories have been developed \cite{Bla05}.
Previous decay simulations were always performed with strong initial
magnetic fields.
Only recently, the need for studying the evolution of hydromagnetic
turbulence in {\em kinetically dominated} systems has been emphasized
\cite{Par17}.
However, no detailed study has been presented as yet, except for our own
work \cite{BKMRPTV17}, which focused on the case without kinetic helicity.

\section{Helical dynamos with time-dependent coefficients}
\label{DynamosCoefficients}

A simple example of a dynamo is one that works owing to the presence
of kinetic helicity, $\bra{\oo\cdot\uu}$, where $\oo=\nab\times\uu$
is the vorticity and $\uu$ is the turbulent velocity.
In stationary isotropic turbulence a statistically averaged mean
magnetic field $\meanBB$ obeys \cite{Mof78,KR80}
\EQ
\partial\meanBB/\partial t=\nab\times\left[
\alpha_{\rm dyn}\meanBB-(\etat+\eta)\mu_0\meanJJ\right],
\EN
where $\alpha_{\rm dyn}\approx-\tau\bra{\oo\cdot\uu}/3$ is the $\alpha$
effect, $\etat\approx\tau\bra{\uu^2}/3$ is the turbulent magnetic
diffusivity, $\eta$ is the microphysical magnetic diffusivity, and
$\meanJJ=\nab\times\meanBB/\mu_0$ is the mean current density
with $\mu_0$ being the vacuum permeability.
If the coefficients are spatially constant and the domain is periodic,
the solutions are eigenfunctions of the curl operator with eigenvalue $k$.
If the coefficients were also constant in time, $|\meanBB|$ would be
proportional to $\exp(\ii\kk\cdot\xx+\gamma t)$.
There would then be a growing solution that obeys
$\gamma=|\alpha_{\rm dyn} k|-\etaT k^2$ with $\etaT=\etat+\eta$ if
$C\equiv|\alpha_{\rm dyn}|/\etaT k_1>1$, where $k_1=2\pi/L$ is the
smallest wave number that fits into the cubic domain of size $L^3$.

We define the fractional helicity $\epsf$ such that
$\bra{\oo\cdot\uu}=\epsf\bra{\uu^2}/\xiK$, where
$\xiK$ is the scale of the energy-carrying eddies,
which we will later identify with the integral scale that is
formally defined in terms of energy spectra.
Thus, $C=\epsf/(\iota k_1\xiK)$, where $\iota=1+3\,\Rm^{-1}$ with
\EQ
\Rm=\urms\xiK/\eta
\EN
being the magnetic Reynolds number, $\tau=\xiK/\urms$ is the turnover time,
and $\urms=\bra{\uu^2}^{1/2}$ is the rms velocity \cite{BB02}.
The effective wave number of the large-scale field, $\km$, is not normally
at the minimal wave number $k=k_1$, but at a larger value, so
$k_1\leq\km\leq(2\xiM)^{-1}$; see, e.g., Fig.~17 of Ref.~\cite{Bra01}.

In decaying hydrodynamic turbulence, we have $\urms^2\propto t^{-p}$ with exponent
$p=10/7$ if the Loitsiansky integral \cite{BP56} is conserved,
or $p=6/5$ if the Saffman integral \cite{Saf67} is conserved.
In these cases, we have
$|\meanBB|=B_0\exp[\int_0^t\gamma(t')\,\dd t']$, where
\EQ
\gamma(t)=(\epsf-\iota\km\xiK)\urms\km/3
\label{gammat}
\EN
with $\epsf=\epsf(t)$, $\xiK=\xiK(t)$, $\km(t)\ge k_1$, and
$\iota=\iota(t)$ now all being time-dependent functions.
Thus, we expect a time-dependent (instantaneous) growth rate
that is, to leading order, given by $\urms(t)\km(t)/3$.
With these preliminary expectations in mind, let us now turn to
three-dimensional turbulence simulations.

\section{Dynamos in decaying turbulence}

We are primarily interested in subsonic turbulence with initial
Mach numbers of the order of 0.1.
At those low Mach numbers, the equation of state no longer affects
the flow (see Fig.~2 of the supplemental material to Ref.~\cite{BK17})
and compressibility effects are unimportant.
We choose to solve for an isothermal gas where the pressure $p$ is
proportional to the local density $\rho$ with $p=\rho\cs^2$.
This equation of state applies to the early universe where
$\cs^2=c^2/3$ with $c$ being the speed of light.
Solving for a weakly compressible gas is computationally
more efficient than solving for an incompressible fluid
where the pressure is a nonlocal function of the velocity.

We neglect kinetic and two-fluid effects in our present work,
which is appropriate for many astrophysical plasmas, including the
early universe \cite{Sub16}.
We thus solve the three-dimensional hydromagnetic equations
\EQ
{\partial\uu\over\partial t}=-\uu\cdot\nab\uu-\cs^2\nab\ln\rho
+{1\over\rho}\left(\JJ\times\BB+\nab\cdot2\rho\nu\SSSS\right),
\label{duu}
\EN
\EQ
{\partial\ln\rho\over\partial t}=-\uu\cdot\nab\ln\rho-\nab\cdot\uu,
\label{dlnrho}
\EN
\EQ
{\partial\AAA\over\partial t}=\uu\times\BB-\eta\mu_0\JJ,
\label{indEq}
\EN
where
${\sf S}_{ij}={1\over2}(u_{i,j}+u_{j,i})-{1\over3}\delta_{ij}\nab\cdot\uu$
is the traceless rate of strain tensor, $\nu$ is the kinematic viscosity,
$\BB=\nab\times\AAA$ is the magnetic field, $\JJ=\nab\times\BB/\mu_0$
is the current density, and $\mu_0$ is the vacuum permeability.
We consider a triply periodic domain of size $L^3$, so the smallest
wave number in the domain is $k_1=2\pi/L$.

We take the initial velocity to be solenoidal and define it
in Fourier space as
\begin{equation}
u_i({\kk})=\left[{\sf P}_{ij}(\kk)+\ii\sigmaK\epsilon_{ijl} {k_l\over k}\right]
{u_0 k_0^{-3/2} g_j({\kk})\, (k/k_0)^{\alpha/2-1}\over[1+(k/k_0)^{2(\alpha+5/3)}]^{1/4}},
\end{equation}
where ${\sf P}_{ij}=\delta_{ij}-k_ik_j/k^2$ is the projection operator,
$\ggg(\kk)$ is the Fourier transform of a spatially $\delta$-correlated
vector field in three dimensions with Gaussian distributed fluctuations,
and $k_0$ is wave number of the peak of the initial spectrum.
It corresponds to the initial wave number of the energy-carrying eddies.
We choose $k_0/k_1=60$.
The exponent $\alpha$ (not to be confused with the mean-field dynamo
coefficient $\alpha_{\rm dyn}$) denotes the slope of the spectrum at
low wave numbers.
We choose $\alpha=4$ for a causally generated solenoidal field \cite{MY71,DC03}.
The fractional initial helicity is controlled by the parameter $\sigmaK$
and given by $\epsf=2\sigmaK/(1+\sigmaK^2)$.
For the initial magnetic field, we take the same spectrum, but
with $\sigmaM$ instead of $\sigmaK$, and amplitude $B_0$ instead of $u_0$.
The velocity is initially fully helical ($\sigmaK=1$) and solenoidal.
We consider an initial $\BB(\kk)$ with $\sigmaM=0$, $1$, and $-1$.
The initial density is constant and given by $\rho_0$.

Viscosity $\nu$ and magnetic diffusivity $\eta$ are usually very small
in physical systems of interest.
This is generally difficult to simulate, especially at early times
if we fix $\nu$ and $\eta$ to be that small.
However, a self-similar evolution is made possible by allowing $\nu$ and
$\eta$ to be time-dependent (after some time $t>t_0$; see below) with
\EQ
\nu(t)=\nu_0\max(t,t_0)^r,
\EN
where $r=(1-\alpha)/(3+\alpha)$ \cite{Ole97}, which gives $r=-3/7$
for $\alpha=4$.
The time $t_0$ is chosen to be short ($t_0 \urms/\xiM\approx1$), but
non-vanishing to prevent $\nu$ and $\eta$ from becoming singular for $r<0$.
In most of the cases reported below, we assume $\eta(t)=\nu(t)/\Pm$,
where $\Pm=1$ is chosen for the magnetic Prandtl number.
In some cases, we also compare with cases where $\Pm\neq1$ and with cases
where $\nu\equiv\nu_0$ and $\eta\equiv\eta_0$ are constant in time.

We define kinetic and magnetic energy spectra, $\EK(k,t)$ and $\EM(k,t)$,
respectively.
They are normalized such that $\int E_i(k,t)\,\dd k={\cal E}_i$ for
$i={\rm K}$ or ${\rm M}$, where $\EEK=\rho_0\urms^2/2$ and $\EEM=\Brms^2/2\mu_0$
are the kinetic and magnetic mean energy densities, and
$\Brms$ is the rms magnetic field.
Time is given in units of the initial turnover time,
$\tau_0=\tau(0)$, and
\EQ
\xi_i(t)=\int_0^\infty k^{-1} E_i(k,t)\,\dd k/{\cal E}_i(t)
\EN
is the integral scale.
We have chosen $t_0/\tau_0=0.1$ for the time when viscosity
and magnetic diffusivity become time-dependent.
Our runs are given in \Tab{list}, where the initial Alfv\'en speed
$\vAz=B_0/\sqrt{\mu_0\rho_0}$ has been introduced and the end time of
the run $t_{\rm e}$ is given.

%A:  150  351   45  20
%B:   60   80
%C:  170  520
%D:   27   15
%E:   27    5
%F:   22   18   40  10
%G:   17    6   20   5

\begin{table}[b!]\caption{Summary of the runs discussed in this paper.
}\vspace{12pt}\centerline{\begin{tabular}{crccccccccc}
Run & $\sigmaK$ & $\sigma_{\rm M}$ & $\vAz/u_0$ & $\Rm$ & $\Lu$ & $t_{\rm e}/\tau_0$ &
$\qM(t_{\rm e})$ & $\pM(t_{\rm e})$ \\
\hline
A & 1 & 0  & $0.1$ &38--323&17--830&49,000& 0.55 & 0.58 \\%KH1152tnuk4b_sig1b_M01
B & 1 & 1  & $0.1$ &35--120&14--182&23,000& 0.46 & 0.59 \\%KH1152tnuk4b_sig1_M01
C & 1 &$-1$& $0.1$ &37--326&29--1090&15,000& 0.53 & 0.57 \\%KH1152tnuk4b_sig1c_M01 (goes to beta=0)
D & 1 & 0  & $0.01$&34--37&3--21& 2500 & 0.38 & 1.10 \\%KH1152tnuk4b_sig1b_M001 (goes to beta=4)
E & 1 & 0  &$0.001$&30--22&0.5--4.5&  400 & 0.35 & 0.70 \\%KH1152tnuk4b_sig1b_M0001
E'& 0 & 0  &$0.001$&30--22&0.5--2.5& 1300 & 0.29 & 0.44 \\%KH1152tnuk4b_sig1c_sigK0_M0001
F & 1 & 0  & $0.1$ &27--21&9--17&   63 & 0.47 & 1.31 \\%KH1152tnuk4c_sig1b_M01
G & 1 & 0  & $0.1$ &14&3--5&   26 & 0.29 & 0.50 \\%KH1152tnuk4d_sig1b_M01
H & 1 & 0  & $0.1$ &8&1--6& 3400 & 0.29 & 0.50 \\%KH1152tnuk4e_sig1b_M01
\label{list}\end{tabular}}\end{table}
In the following, we characterize the values of $\nu_0$ and $\eta_0$
by the time-dependent magnetic Reynolds and Lundqvist numbers,
\EQ
\Rm=\urms\xiM/\eta\quad\mbox{and}\quad
\Lu=\Brms\xiM/\eta,
\EN
respectively.
Their initial and final values are indicated in \Tab{list},
respectively.
Note also that we have now chosen to define $\Rey$ and $\Rm$ in terms
of $\xiM$ instead of $\xiK$.
We do this because the magnetic energy spectrum has a more clearly
defined peak, while that of the kinetic energy spectrum is less clear and,
at least after some time, its evolution is enslaved by the magnetic field.
Furthermore, we define instantaneous scaling exponents of ${\cal E}_i(t)$
and $\xi_i(t)$ as
\EQ
p_i(t)=d\ln{\cal E}_i/d\ln t,\quad
q_i(t)=d\ln\xi_i/d\ln t,
\EN
and plot $p_i(t)$ versus $q_i(t)$
for $i={\rm M}$ and ${\rm K}$ and discuss the evolution of the point
\begin{equation}
\PP_i=(p_i,q_i)
\end{equation}
in the $pq$ diagram. Solutions that obey invariance under rescaling \cite{Ole97,Ole15,BK17},
\begin{equation}
k\to k'\ell^{-1}\quad\mbox{and}\quad
t\to t'\ell^{1/q_i},
\end{equation}
all lie on the line $p_i=2(1-q_i)$ in this diagram.

In the case of a self-similar evolution \cite{Ole97,Ole15}, the magnetic energy spectra
can be described by a single function $\phi(k\xiM)$ of the product
$k\xiM$ such that \cite{BK17}
\EQ
\EM(k\xiM(t),t)\approx\xiM^{-\betaM}\phi(k\xiM),
\label{Collapse}
\EN
where $\phi(k\xiM)$ is a function of magnetic Reynolds and Prandtl numbers,
but not of time.
Note that $\xiM(t)$ varies such that the peak of the spectrum is always
at $k\xiM\approx1$.
If the solutions are invariant under rescaling, they must obey
$q_i=2/(\beta_i+3)$ \cite{Ole97}.

By integrating $\EEM(t)=\int\EM(k,t)\,\dd k$, one can see that
\EQ
\EEM(t)\propto\xiM^{-(\betaM+1)}\propto t^{-\qM(\betaM+1)},
\label{Eq13}
\EN
and therefore we have $1+\betaM=\pM/\qM$ \cite{BK17}.
On dimensional and physical grounds \cite{BM99}, one expects the rate
of change to obey
\EQ
{\dd\EEM\over\dd t}\propto\xiM^{-1}\EEM^{3/2}.
\EN
Further details regarding the relation between $\xiM$ and $\EEM$
depend of the conservation laws that are being obeyed.
For example, when magnetic helicity is conserved, we have
$\bra{\AAA\cdot\BB}\propto\EEM\xiM=\const$, so $\xiM\propto\EEM^{-1}$ and therefore
$\dd\EEM/\dd t\propto\EEM^{5/2}$, which yields \cite{BM99}
\EQ
\pM=\qM=2/3.
\EN
This, in turn, implies $\betaM=0$, so the height of the peak of
$\EM(k,t)$ stays unchanged; see \Eq{Collapse}.

For our numerical simulations we use the {\sc Pencil Code}
(\url{https://github.com/pencil-code}), a public MHD code
that is particularly well suited for simulating turbulence.
In all cases we use $1152^3$ meshpoints, which is large enough to
ensure that the inverse-cascade effects are well reproduced; see
Ref.~\cite{BKT15} for earlier work highlighting the importance of
high resolution in connection with the inverse cascade in nonhelical
hydromagnetic turbulence.

\begin{figure}[t!]\begin{center}
\includegraphics[width=.7\columnwidth]{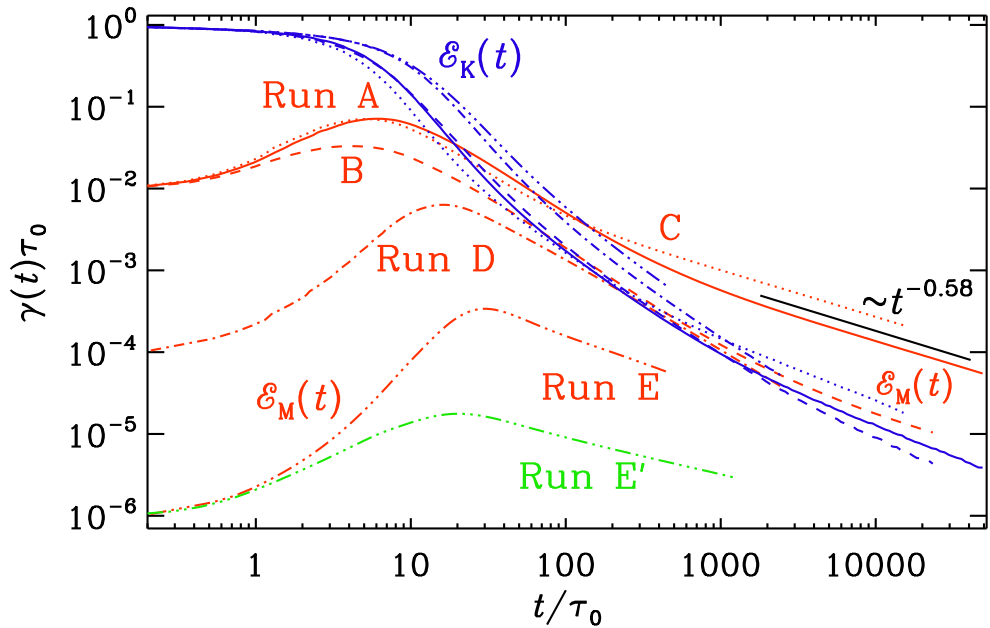}
\end{center}\caption[]{
Evolution of $\EEK$ (blue) and $\EEM$ (red) for
$\sigmaM=0$ (solid), $\sigmaM=1$ (dashed),
and $\sigmaM=-1$ (dotted) for $\vAz/u_0=0.1$ (Runs~A--C),
as well as $0.01$ (dot-dashed, Run~D) and $0.001$ (triple dot-dashed, Run~E)
for $\sigmaM=0$.
The green triple dot-dashed line denotes Run~E',
which has zero initial kinetic helicity.
}\label{pcomp}\end{figure}

\section{Results}

\subsection{Kinetic and magnetic energy evolution}

In \Fig{pcomp}, we plot $\EEK(t)$ and $\EEM(t)$ for Runs~A--E.
$\EEM$ is found to increase at first, reaches a maximum at
$t/\tau_0\approx10$, and then approaches a late-time magnetic decay law
approximately proportional to $t^{-p}$ with $p\ga0.5$.
We see that kinetic energy is transferred to magnetic energy,
whose value eventually exceeds $\EEK$.
The time when this happens depends on the initial magnetic energy.
For Run~A with $\vAz/u_0=0.1$, this time is $t/\tau_0\approx20$;
see \Fig{pcomp}, and for Run~D it is $t/\tau_0=200$.

\begin{figure}[t!]\begin{center}
\includegraphics[width=.7\columnwidth]{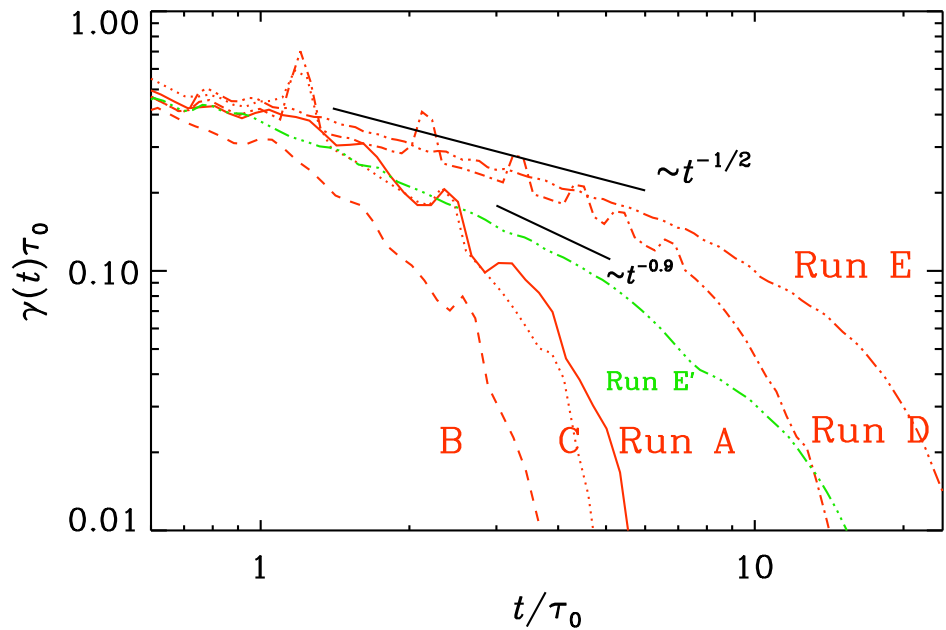}
\end{center}\caption[]{
Instantaneous growth rates $\gamma(t)\tau_0$ of $\Brms$ for
$\sigma_{\rm M}=0$ (solid), $\sigma_{\rm M}=1$ (dashed),
and $\sigma_{\rm M}=-1$ (dotted) for $\vAz/u_0=0.1$ (Runs~A--C),
as well as $0.01$ (dot-dashed, Run~D) and $0.001$ (triple dot-dashed, Run~E).
The green triple dot-dashed line denotes Run~E' with
zero initial kinetic helicity.
}\label{pcomp_linlog}\end{figure}

Although the turbulence is decaying, it is still possible to define a
meaningful growth rate of the magnetic field and to estimate a critical
value of the magnetic Reynolds number above which dynamo action is
possible.
We do this by plotting the instantaneous growth rate,
\EQ
\gamma(t)=\dd\ln\Brms/\dd t,
\EN
of the rms magnetic field $\Brms$.
The result is shown in \Fig{pcomp_linlog}, where we plot $\gamma(t)\tau_0$
versus $t/\tau_0$.
We see that the values with the weakest initial field (i.e., in the
kinematic limit) are $\gamma(t)\tau_0\la0.5$ at early times.
At later times, however, $\gamma(t)$ decreases.
This is roughly consistent with \Eq{gammat}.
Furthermore, the decay is faster if $\eta$ is larger, i.e., $\Rm$ smaller.

In the case with zero kinetic helicity, the initial growth rate is nearly
the same as with kinetic helicity; see Run~E' in \Fig{pcomp_linlog}.
This suggests that there is also small-scale dynamo action.
Owing to the absence of kinetic helicity, $\gamma(t)$ follows a
slightly steeper power law of the approximate form $t^{-0.9}$.
At later times, however, the magnetic field of Run~E' decays in the
same way as that of Run~E.

\begin{figure}[t!]\begin{center}
\includegraphics[width=.7\columnwidth]{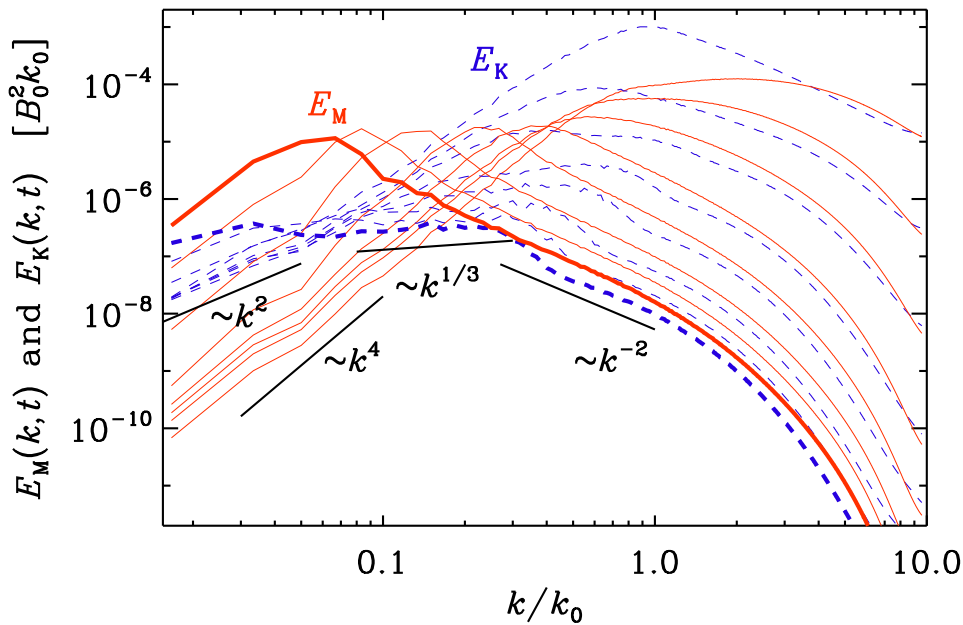}
\end{center}\caption[]{
$\EK(k,t)$ and $\EM(k,t)$ for $t/\tau=16$, $60$, $200$, $800$, $2000$,
$8000$, and $17,000$, for Run~A.
Time is decreasing downward and the last time is shown as fat lines.
}\label{pkt1152_KH1152tnuk4b_sig1b_M01}\end{figure}

To understand what has happened, we look at the spectra $\EK(k,t)$ and
$\EM(k,t)$ in \Fig{pkt1152_KH1152tnuk4b_sig1b_M01}.
We see that, during the late evolution ($t/\tau_0>1000$), the magnetic energy
spectra are shape-invariant and just translate toward smaller $k$.
This is suggestive of an inverse cascade, where
$\EM(k\xiM(t),t)$ collapses onto the same curve $\phi(k\xiM)$ with
$1+\betaM=\pM/\qM$ \cite{BK17}; see \Eq{Collapse}.
Here the correlation length $\xiM$ increases like $t^\qM$ such that
$\bra{\BB^2}\xiM^{1+\betaM}$ stays constant; see \Eq{Eq13}.
The value of this constant depends on the total amount of magnetic helicity
that is {\em produced} in the system.
To compensate for the decay in magnetic energy, we multiply $\EM$
by $\xiM^{\betaM}$ with an exponent $\betaM$ such that the
compensated spectra collapse onto a single function

\begin{figure}[t!]\begin{center}
\includegraphics[width=.7\columnwidth]{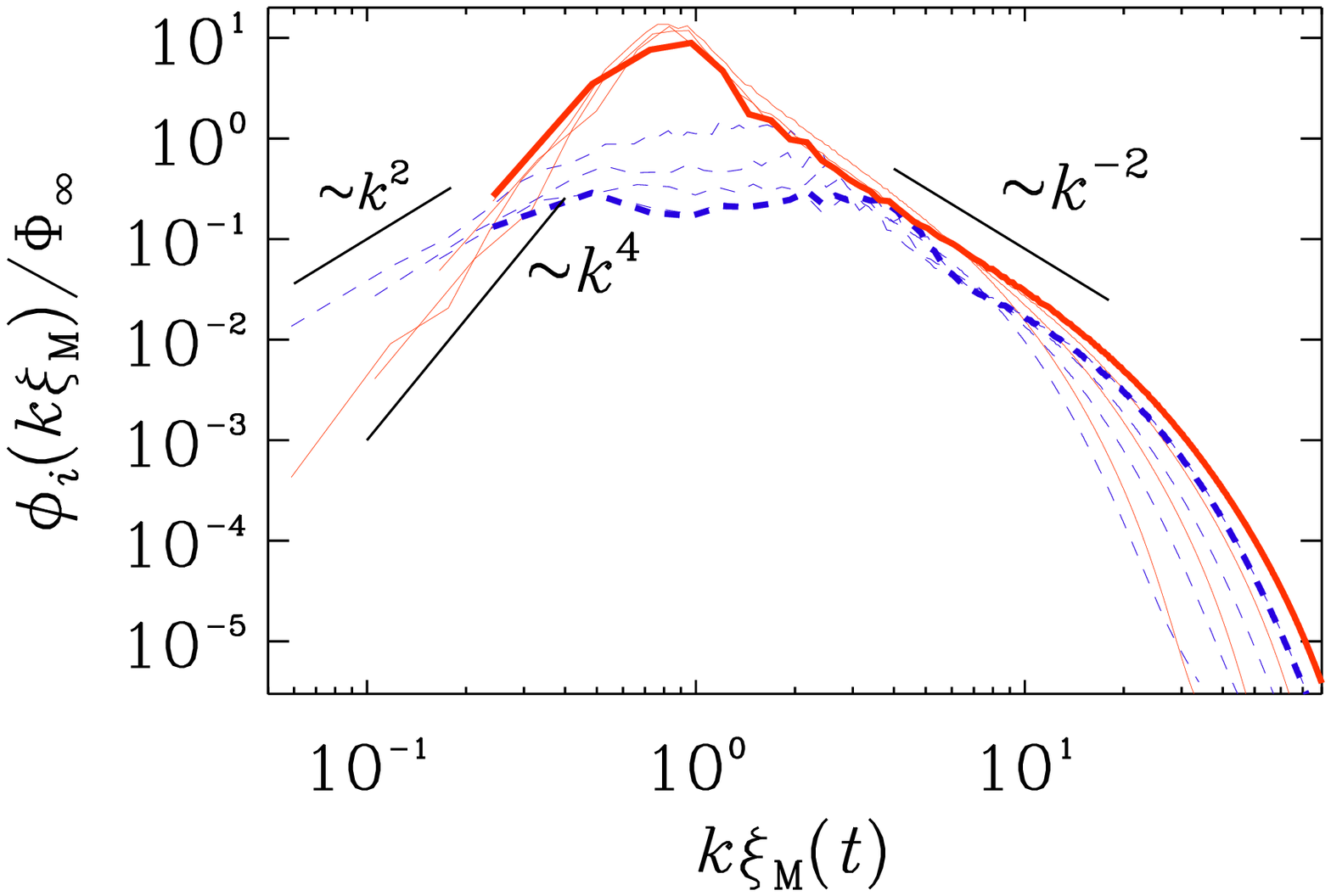}
\end{center}\caption[]{
$\EK(k,t)$ and $\EM(k,t)$ at $t/\tau=2600$, $9400$, $24,000$, and $45,000$,
collapsed spectra using $\betaM=0$
for Run~A.
}\label{pkt1152_kalp_one_KH1152tnuk4b_sig1b_M01}\end{figure}

\begin{figure}[t!]\begin{center}
\includegraphics[width=.7\columnwidth]{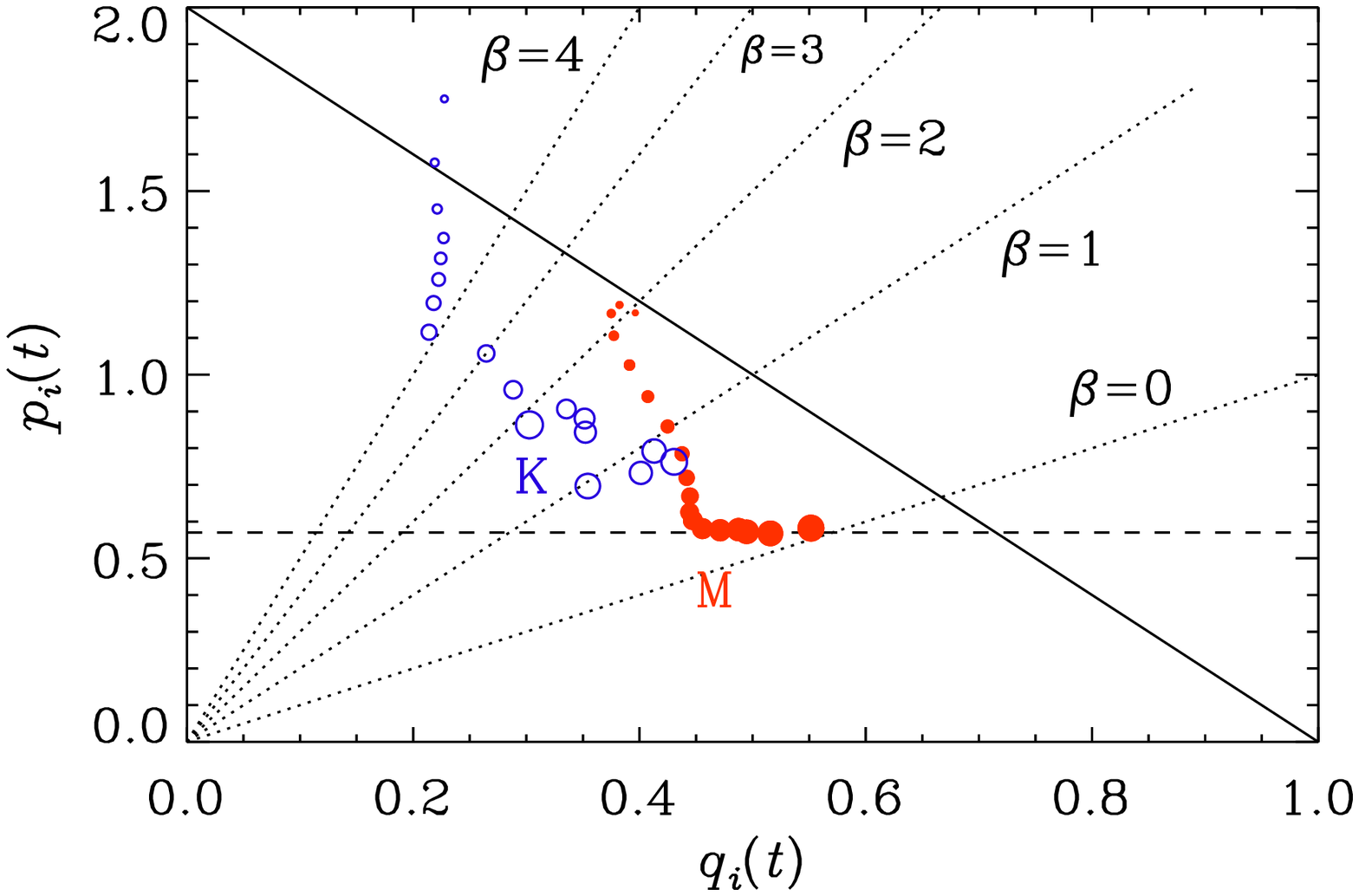}
\end{center}\caption[]{
$pq$ diagram for Run~A for kinetic (blue open symbols) and
magnetic (red filled symbols) energy spectra.
Near the end of the run (larger symbols), the solution evolves
along the $\pM=0.58$ line (dashed) and $\qM\approx0.55$ with
$\betaM=\pM/\qM-1\approx0.05$ is found at the end of the run.
Smaller (larger) symbols denote earlier (later) times.
}\label{pq_KH1152tnuk4b_sig1b_M01}\end{figure}

\begin{figure}[t!]\begin{center}
\includegraphics[width=.7\columnwidth]{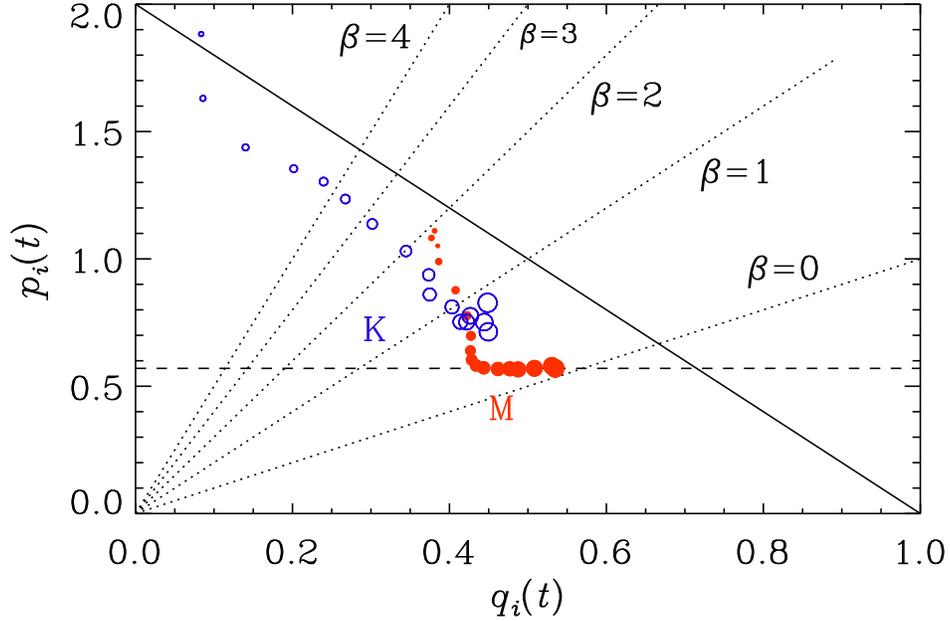}
\end{center}\caption[]{
Same as \Fig{pq_KH1152tnuk4b_sig1b_M01}, but for Run~C, where the solution
evolves along the $\pM=0.57$ line (dashed) and $\qM\approx0.53$ with
$\betaM=\pM/\qM-1\approx0.08$ is found at the end of the run.
}\label{pq_KH1152tnuk4b_sig1c_M01}\end{figure}

A closer inspection of the magnetic decay gives $\qM\approx0.55$ and
$\pM\approx0.58$ at the end time for Run~A, so that $\betaM\approx0.05$;
see the $pq$ diagram in \Fig{pq_KH1152tnuk4b_sig1b_M01}.
In \Fig{pq_KH1152tnuk4b_sig1c_M01} we show a similar plot for Run~C.
Since the magnetic decay is not truly self-similar, it does not obey the
scaling relation $\betaM=2/\qM-3$ \cite{Ole97} and does not fall on
the line $\pM=2\,(1-\qM)$, which is indicated in
\Fig{pq_KH1152tnuk4b_sig1b_M01} by a solid line.

\begin{figure}[t!]\begin{center}
\includegraphics[width=.7\columnwidth]{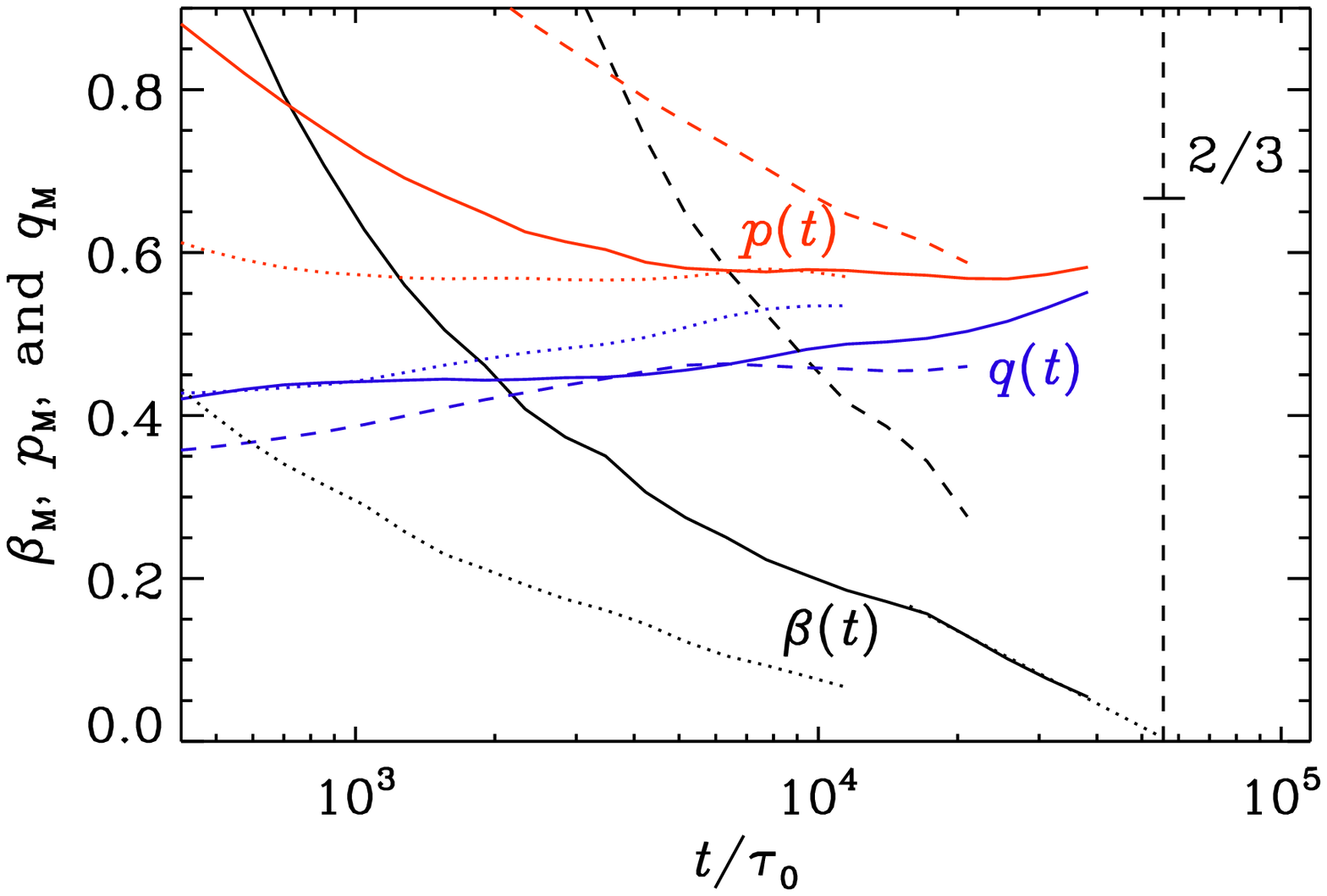}
\end{center}\caption[]{
Evolution of $\betaM(t)$ (black), $\pM(t)$ (red), and $\qM(t)$ (blue) for Run~A
(solid lines), Run~B (dotted lines), and Run~C (dashed line).
Note that $\betaM(t)$ would reach zero at an extrapolated time
of $t_\ast\approx5300$.
}\label{pq_time_KH1152tnuk4b_sig1b_M01}\end{figure}

At late times, although $\pM$ and $\qM$ are still different from the expected
law with $\pM=\qM=2/3$, there are several other similarities to earlier
calculations of magnetically dominated hydromagnetic turbulence.
In particular, we see a change of the low wave number slope of $\EK$
from $k^4$ to $k^2$ at later times and at small $k$.
This is a consequence of compressibility \citep{KTBN13,BKT15} and is
not seen in the incompressible case; see the supplemental material of
Ref.~\cite{BL14}.
At larger wave numbers, near the point where $\EM$ peaks,
the kinetic energy is proportional to $k^{1/3}$;
see \Figs{pq_KH1152tnuk4b_sig1b_M01}{pq_KH1152tnuk4b_sig1c_M01} and
Ref.~\cite{BKT15}.
The $k^2$ law for the kinetic energy $\EEK$ is likely a consequence of
turbulent interactions over the scale of the domain since the initial time.

To inspect the slow changes of $\betaM(t)$, $\pM(t)$, and $\qM(t)$ in more
detail, we show in \Fig{pq_time_KH1152tnuk4b_sig1b_M01} their evolution
for Run~A using again a logarithmic time axis.
We see that there is an intermediate plateau when their values are indeed
approximately constant.
At late times, however, we see that $\betaM(t)$ is well described by an
expression of the form $\betaM(t)=\betaMz-\betaMo\ln(t/\tau_0)$.
This implies that $\exp(\betaM-\betaMz)=(t/\tau_0)^{-\betaMo}$.
We also see that the extrapolated time $t_\ast$ when $\betaM(t_\ast)=0$
is given by $t_\ast/\tau_0=\exp(\betaMz/\betaMo)$.
Looking at \Fig{pq_time_KH1152tnuk4b_sig1b_M01} suggests that the
exponents $\pM$ and $\qM$ both increase, although it is not obvious that
they attain the value $2/3$ by the extrapolated time $t_\ast\approx5300$.

\subsection{Effect of finite initial magnetic helicity}

\begin{figure}[t!]\begin{center}
\includegraphics[width=.7\columnwidth]{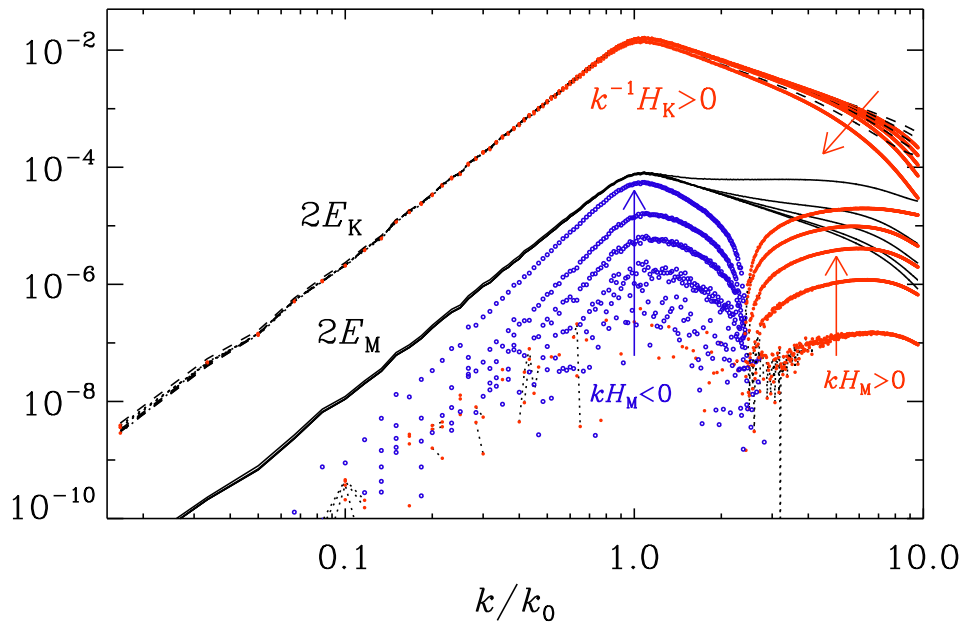}
\end{center}\caption[]{
$k^{-1}\HK(k,t)$ (red) and $k\HM(k,t)$ along with
$2\EK(k,t)$ and $2\EM(k,t)$ (black lines)
at $t/\tau=0.05$, $0.16$, $0.3$, $0.6$, and $1.7$ for Run~A
(red for positive values and blue for negative values).
The blue and red arrows indicate the change of $k\HM(k,t)$
with time.
}\label{pkt1152_panels_KH1152tnuk4b_sig1b_M01_rep}\end{figure}

We recall that, except for Runs~B and C, no magnetic helicity was present
initially, i.e., $\sigma_{\rm M}=0$; see \Tab{list}.
Magnetic helicity is a conserved quantity and it can
only change through resistive effects and at small scales.
To understand how magnetic helicity gets produced, we show in
\Fig{pkt1152_panels_KH1152tnuk4b_sig1b_M01_rep} magnetic and
kinetic helicity spectra, $\HK(k,t)$ and $\HM(k,t)$, respectively.
They obey the realizability conditions, $k^{-1}|\HK(k,t)|\leq2\EK(k,t)$ and
$k|\HM(k,t)|\leq2\EM(k,t)$, respectively, and are normalized such that
$\int\HK\,\dd k=\bra{\oo\cdot\uu}$ and
$\int\HM\,\dd k=\bra{\AAA\cdot\BB}$.

\begin{figure}[t!]\begin{center}
\includegraphics[width=.7\columnwidth]{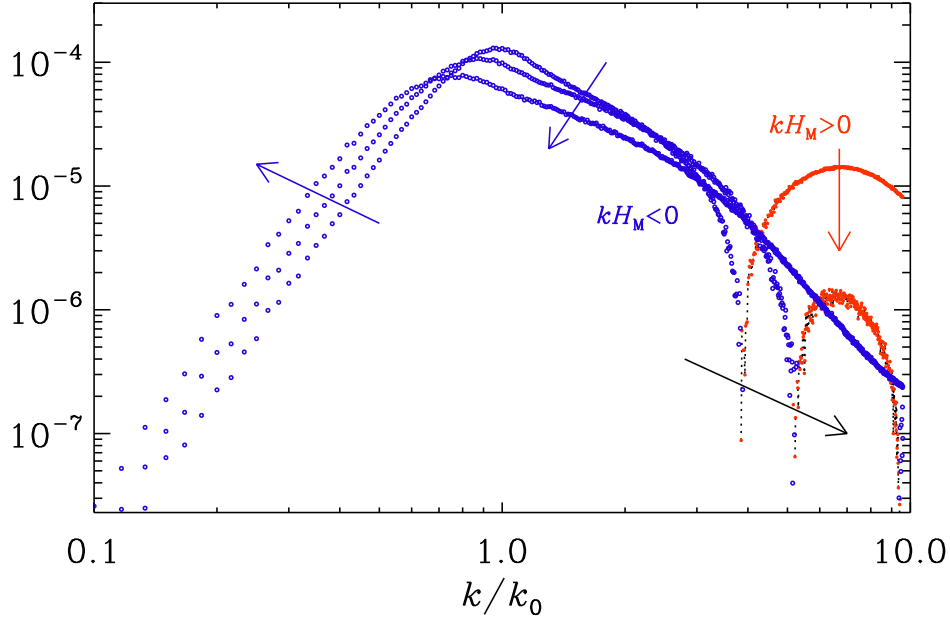}
\end{center}\caption[]{
$k\HM(k,t)$ (red for positive values
and blue for negative values) at later times:
$t/\tau=5$, $10$, and $25$ for Run~A.
The arrows indicate the temporal change of $k\HM(k,t)$.
}\label{pkt1152_panels_late_KH1152tnuk4b_sig1b_M01}\end{figure}

We see that, at early times, a bihelical magnetic helicity spectrum
is produced, where positive and negative contributions are present
simultaneously, though separated in $k$ space, just like in driven
turbulence \cite{Bla05,Bra01}.
Thus, there remains a near-cancelation of the net magnetic helicity
until the magnetic helicity spectrum saturates at $k=O(k_0)$.
When that happens, magnetic helicity at large scales continues to increase
only slowly such that at small scales magnetic helicity continues
to dissipate resistively.
Eventually, at late times, the positive magnetic helicity at small scales has
disappeared and it has at all wave numbers a negative sign; see
\Fig{pkt1152_panels_late_KH1152tnuk4b_sig1b_M01}.
Since Run~C starts with $\sigma_{\rm M}=-1$, we do not need to wait until
the field with positive magnetic helicity gets dissipated.
This leads to a more efficient transfer of kinetic energy to magnetic
energy, which is why we see a stronger growth in \Fig{pcomp}.
The opposite happens in the case with $\sigma_{\rm M}=1$, where the
entire spectrum has initially the `wrong' (positive) sign, making it
even harder to establish a negative magnetic helicity at all wave numbers.
The total magnetic energy decays then subject to resistive decay in the
presence of magnetic helicity.

\subsection{Interpretation}

\begin{figure}[t!]\begin{center}
\includegraphics[width=.7\columnwidth]{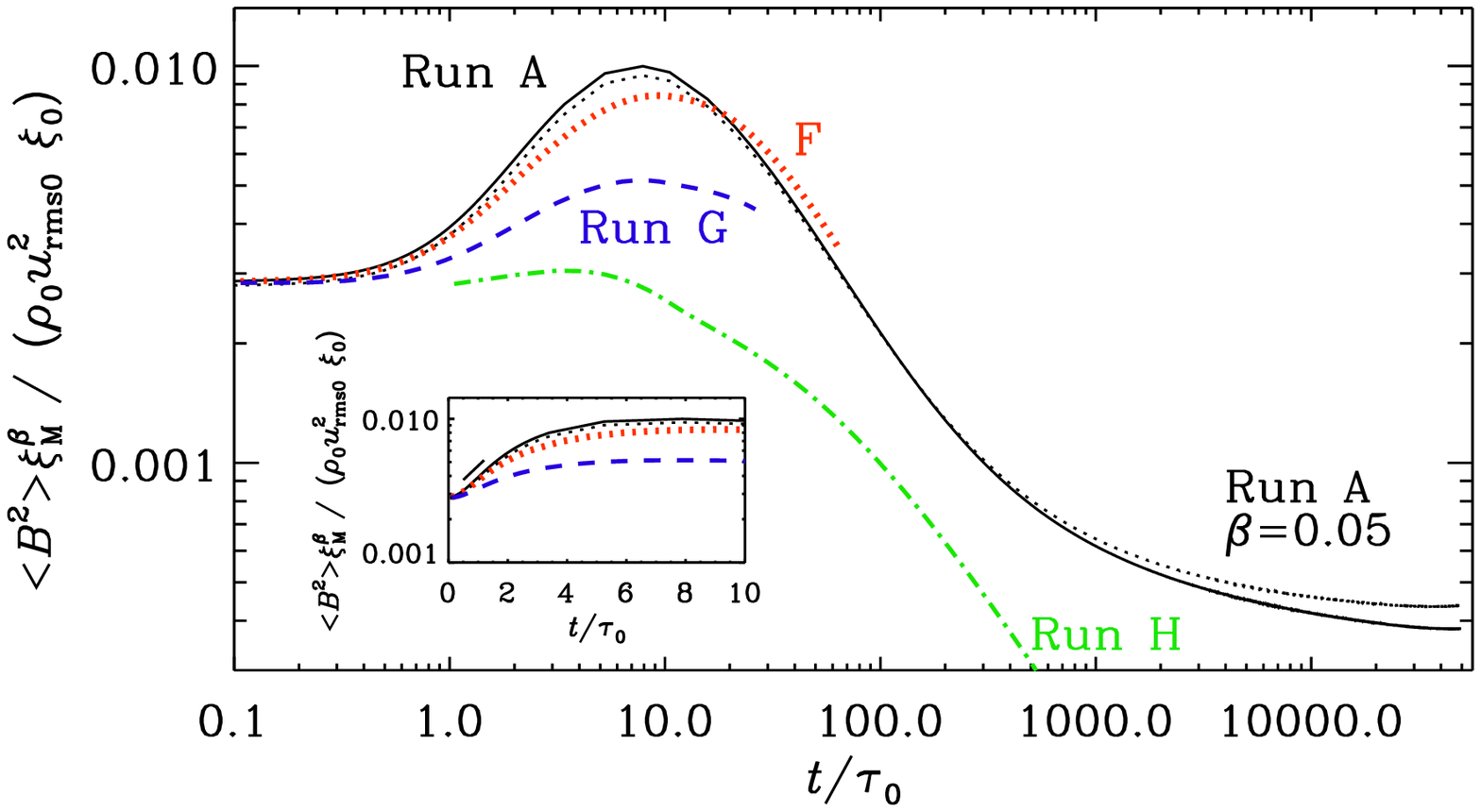}
\end{center}\caption[]{
Evolution of $\bra{\BB^2}\xiM^\beta$ with $\beta=0$ for Run~A (black
solid), F (red dotted), G (blue dashed), and H (green dash-dotted).
In fully helical turbulence, we expect $\bra{\BB^2}\xiM\to\const$,
but here $\bra{\BB^2}\xiM^{1+\betaM}\approx\const$ with $\betaM=0.05$
at the end of the run.
}\label{ppkft}\end{figure}

In \Fig{ppkft} we plot the evolution of $\bra{\BB^2}\,\xiM$ for different
Reynolds numbers (Runs~A, F, and G).
We see that the magnetic helicity produced depends on the magnetic
Reynolds number.
For comparison, we also plot $\bra{\BB^2}\,\xiM^{1.05}$ for Run~A.
This is indicated by the dotted line, which has a flat tangent at the
last time, and corresponds to $\betaM=0.05$.

\begin{figure}[t!]\begin{center}
\includegraphics[width=.6\columnwidth]{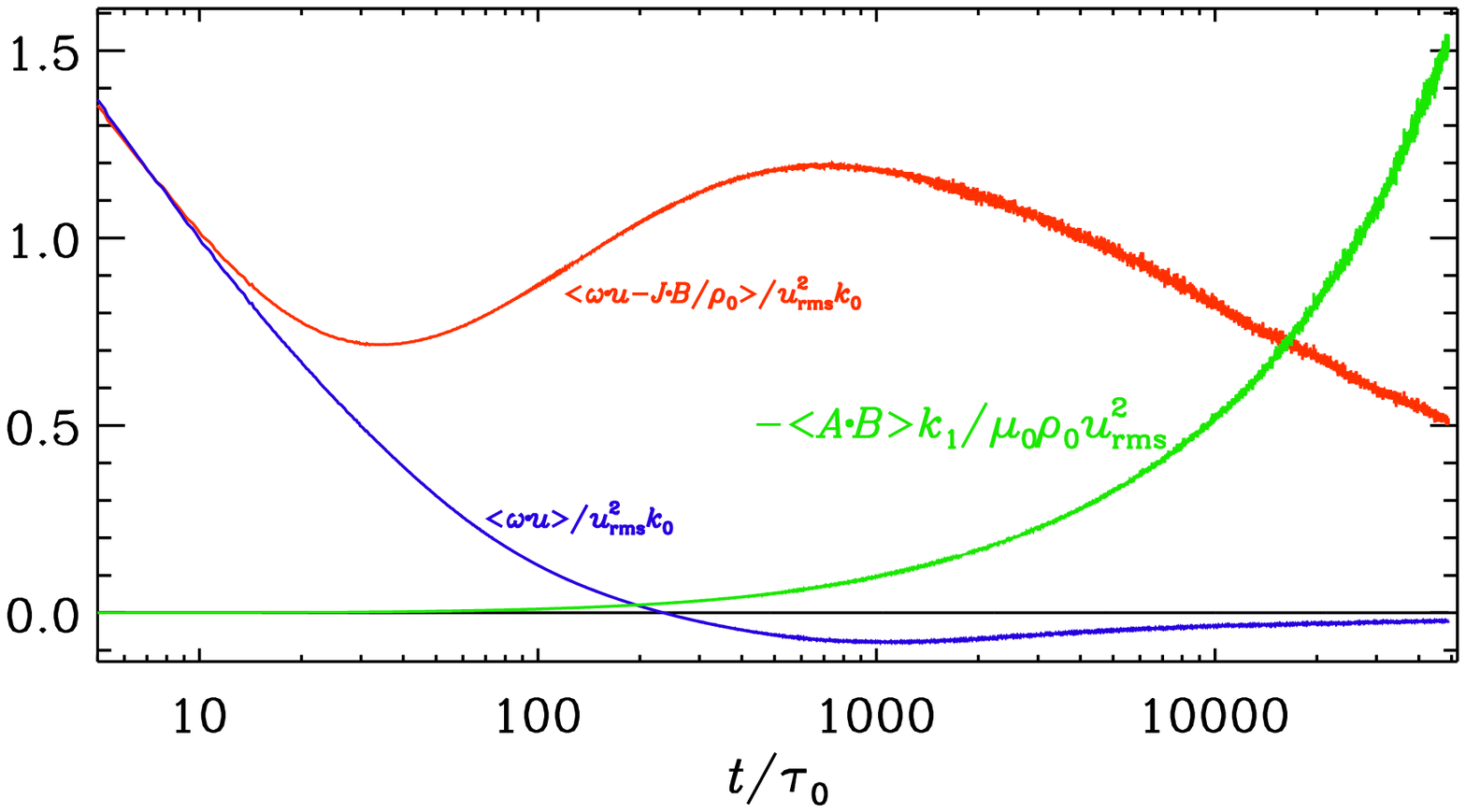}
\end{center}\caption[]{
Evolution of $\bra{\oo\cdot\uu}$ (blue),
$\bra{\oo\cdot\uu-\JJ\cdot\BB/\rho_0}$ (red),
and $\bra{\AAA\cdot\BB}$ (green) for $\Rey=160$ (Run~A).
}\label{phel}\end{figure}

To make contact with the mean-field interpretation developed
in \Sec{DynamosCoefficients}, we show in \Fig{phel}
that $\bra{\oo\cdot\uu}$ dies out while $-\bra{\JJ\cdot\BB}$ increases
such that $\bra{\oo\cdot\uu}-\bra{\JJ\cdot\BB}/\rho_0\approx\const$
during the first 10,000 turnover times.
It is this combination of kinetic and current helicity densities that
replaces the otherwise kinematic $\alpha$ effect in the nonlinear regime
\cite{PFL76,Bla05}.
At $t/\tau_0\approx200$, the sign of $\bra{\oo\cdot\uu}$ changes and
has now the same sign as $\bra{\JJ\cdot\BB}$.
This can be explained by the strong dominance of the magnetic field
over the velocity field, which begins already at $t/\tau_0\approx20$;
see \Fig{pcomp}.

In \Fig{phel}, we also plot $\bra{\AAA\cdot\BB}$ and see that it never
reaches a constant---not even until the end of the run.
This explains why $\pM$ and $\qM$ are still different from $2/3$.
By comparison with the other helicities, the magnetic helicity appears to
rise sharply at $t/\tau_0\ga3\times10^4$ in this double-logarithmic plot.
This signals the end of the $\pM\approx1/2$ scaling of magnetic energy
and the beginning of a $t^{-2/3}$ scaling after even later times.

\begin{figure}[t!]\begin{center}
\includegraphics[width=\columnwidth]{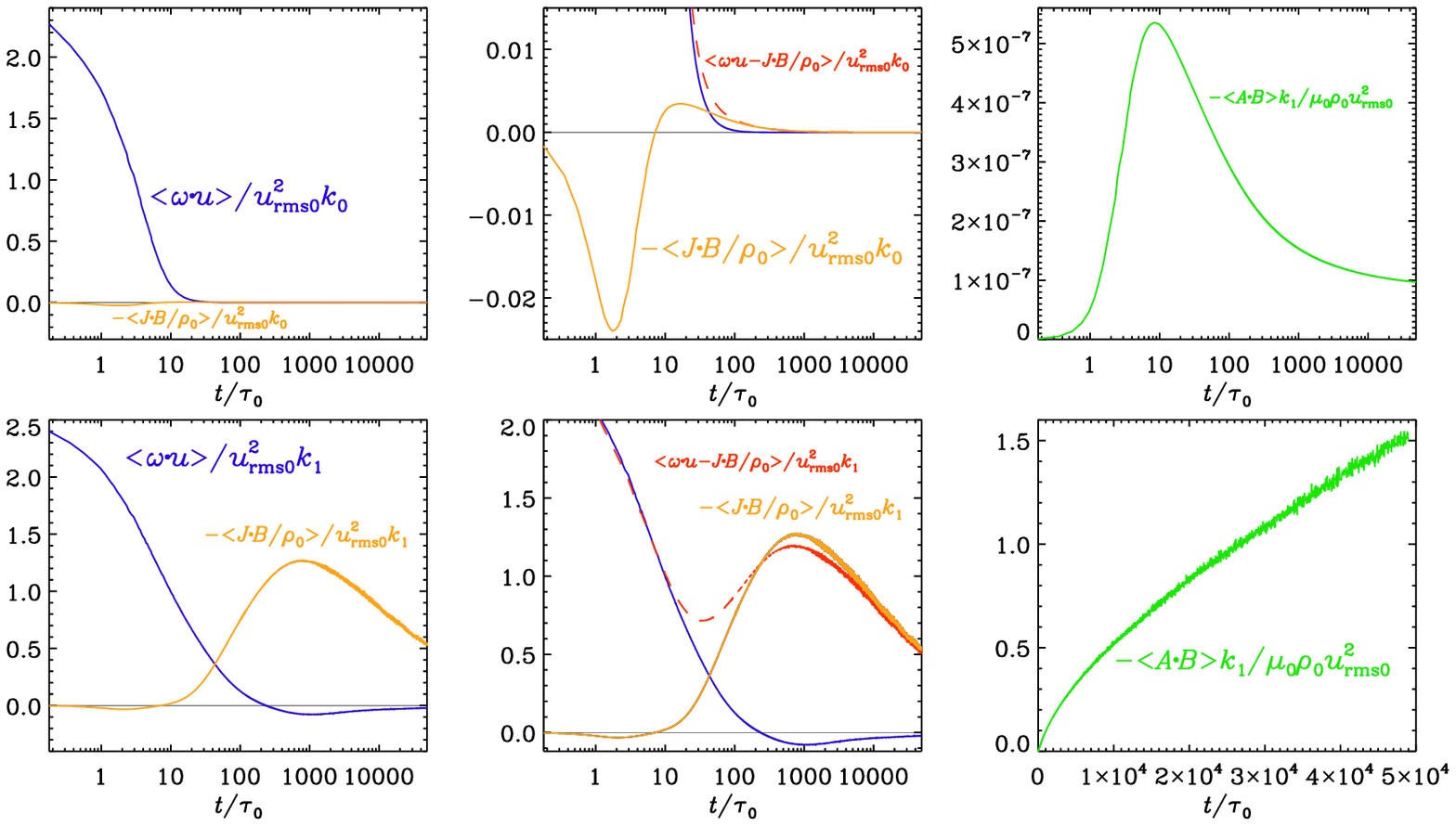}
\end{center}\caption[]{
Similar to \Fig{phel}, but with different normalizations
and in separate panels for
$\bra{\oo\cdot\uu}$ (blue),
$\bra{\oo\cdot\uu-\JJ\cdot\BB/\rho_0}$ (red),
$-\bra{-\JJ\cdot\BB/\rho_0}$ (orange),
and $\bra{\AAA\cdot\BB}$ (green) for $\Rey=160$ (Run~A).
}\label{phel3}\end{figure}

We emphasize that in \Fig{phel} we have plotted the time axis
logarithmically and have normalized by the time-varying rms velocity.
In this way we were able to display the various sign changes of
kinetic and current helicities, but it also distorted the view.
For this reason we now show in \Fig{phel3} in separate panels the
kinetic, current, and magnetic helicities with a constant normalization
using the initial velocity and the initial peak wavenumber
along with time-dependent normalizations using the wavenumber
of the domain, $k_1$, and a linear time axis for the magnetic helicity.
We see that magnetic helicity is always negative, reaches a peak
at $t/\tau_0=10$ and then decays, before asymptoting to a finite value.
When normalized by $\urms^2$, the modulus of the magnetic energy
increases approximately linearly.

The current helicity normalized by $\urms^2$ shows a negative peak at
$t/\tau_0=1000$.
This is when $\bra{\oo\cdot\uu}$ reached a negative peak,
confirming again that the reason for its sign change is indeed
related to the current helicity, which is then also negative and
much stronger than the kinetic helicity.

\subsection{Robustness of the $t^{1/2}$ scaling}

\begin{figure}[t!]\begin{center}
\includegraphics[width=.49\columnwidth]{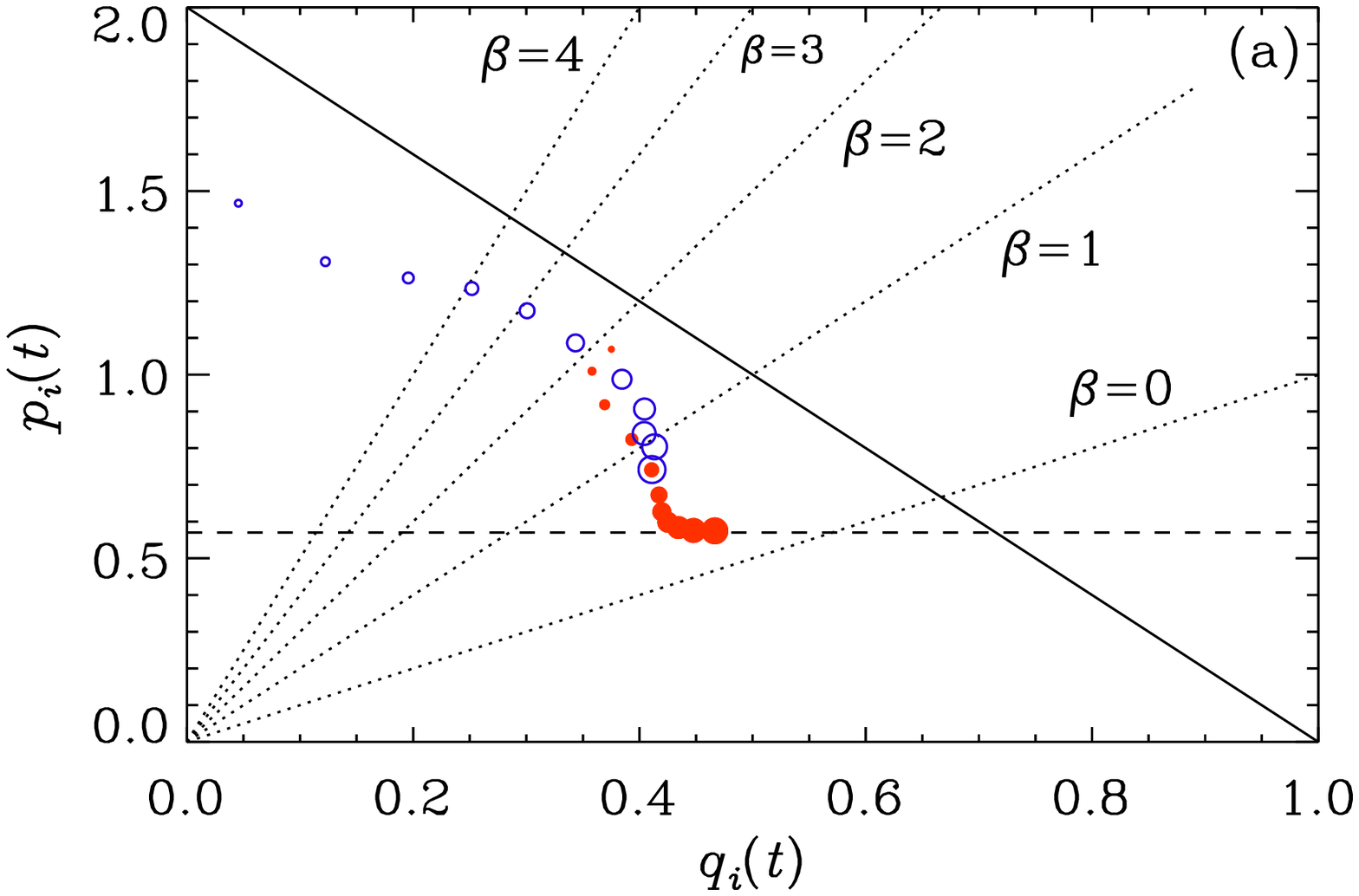}
\includegraphics[width=.49\columnwidth]{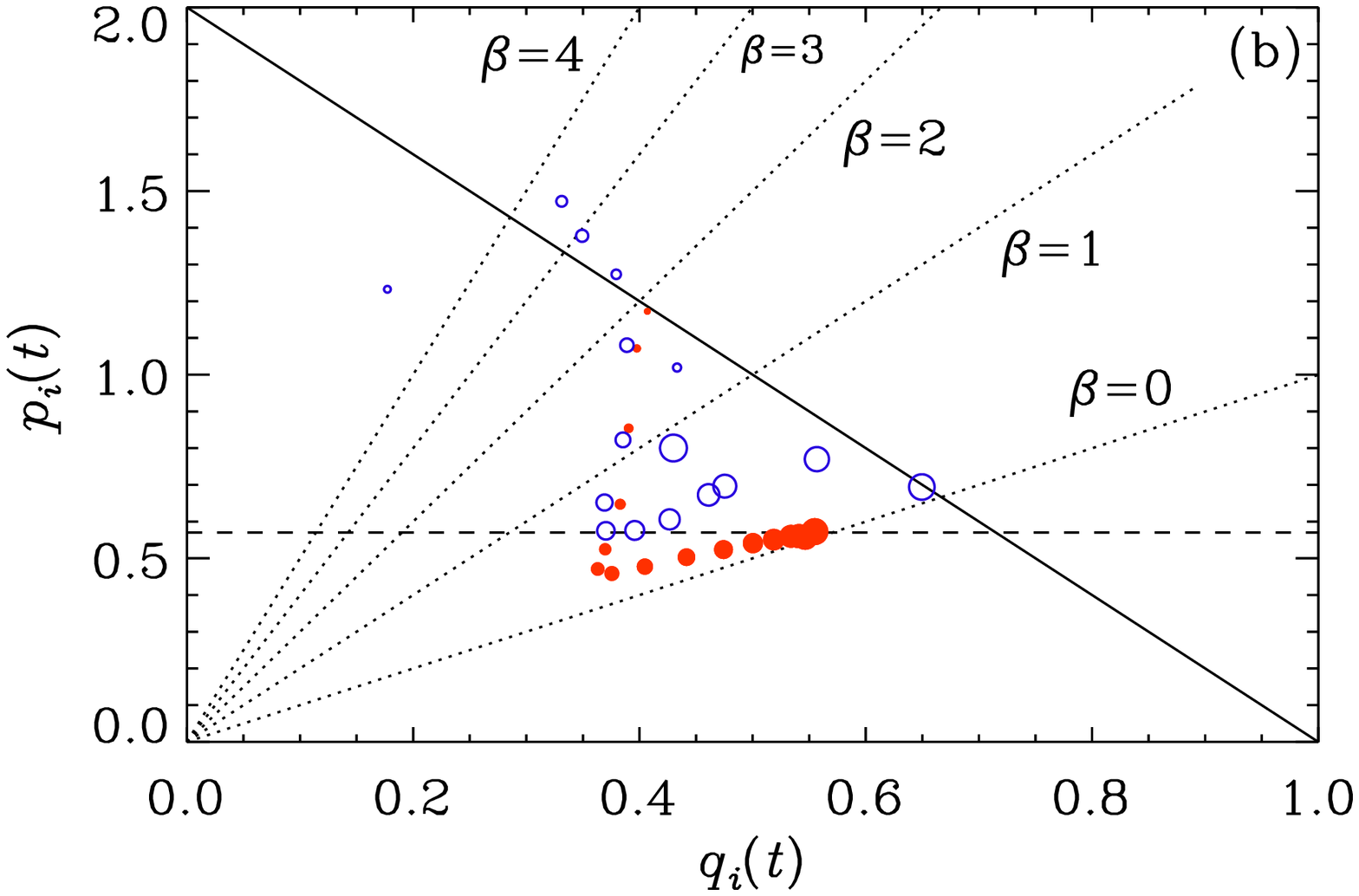}
\end{center}\caption[]{
$pq$ diagram for $\Pm=0.1$ (a) and $10$ (b).
Smaller (larger) symbols denote earlier (later) times.
The $\pM=0.58$ line (dashed) is shown for comparison.
}\label{pq_KH1152tnuk4b_sig1c_M01_Pr}\end{figure}

It is here for the first time that the $t^{1/2}$ scaling has been observed.
Several potentially important assumptions have been made and it needs
to be seen to what extent they might affect our findings.
Here we examine both the assumption of using $\Pm=1$ and the assumption
of using a time-dependent viscosity and a time-dependent magnetic diffusivity.
In \Figsp{pq_KH1152tnuk4b_sig1c_M01_Pr}{a}{b}, we plot the results for
$\Pm=0.1$ and $10$, respectively.
In both cases, a similar evolution of $(\pM,\qM)$ along the line $\pM\approx1/2$
is seen while $\qM$ increases and approaches the $\pM=2(1-\qM)$ line.
Reaching this point would require a larger dynamical range and thus much
larger domains and computation times than what has been possible so far.
This is because, in the present runs, $k_1\xiM$ becomes rather small
($\la3$) toward the end of the run, so inverse transfer is no longer
independent of the system size.

\begin{figure}[t!]\begin{center}
\includegraphics[width=.7\columnwidth]{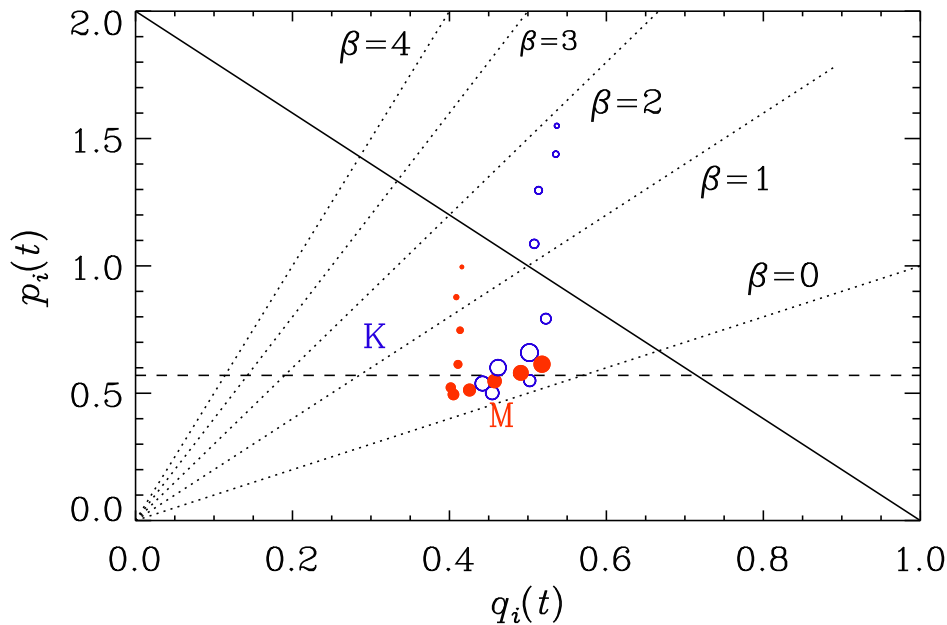}
\end{center}\caption[]{
$pq$ diagram with constant $\nu=\eta=10^{-6}$ for $\Pm=1$.
Again, smaller (larger) symbols denote earlier (later) times
and the $\pM=0.58$ line (dashed) is shown for comparison.
}\label{pq_KH1152k4b_sig1c_M01}\end{figure}

In \Figp{pq_KH1152tnuk4b_sig1c_M01_Pr}{b}, we also see that the trajectory
overshoots the $\pM\approx1/2$ line when $\Pm=10$.
This overshooting indicates that
$\eta$ is still too small for our numerical resolution of $1152^3$ meshpoints.
We have seen a similar behavior also when using a time-independent, but
with a very small value of $\nu=\nu_0$; see \Fig{pq_KH1152k4b_sig1c_M01}
for such an example.

Thus, the principal finding of an evolution along the $\pM\approx1/2$
line with increasing $\qM$ toward the $\beta=0$ line, which is shortly
before it reaches the $\pM=2(1-\qM)$ equilibrium line, is recovered over
a range of different circumstances, but the quality of convergence depends
on how well we can approach the high magnetic Reynolds number limit.

\section{Conclusions}

Our work has demonstrated for the first time that the decay of turbulence
with kinetic helicity leads to a nonconventional intermediate magnetic decay law
with $\pM\approx1/2$ and $\qM$ slowly increasing from about 0.4 to 0.6 before both
$\pM$ and $\qM$ are expected to approach 2/3.
Qualitatively, our results are easily explained.
At early times, a bihelical magnetic helicity spectrum develops and
it grows until it reaches equipartition at the wave number where the
magnetic spectrum peaks.
At small scales, the sign of the magnetic helicity agrees with that of the
kinetic helicity.
At early times, however, no net magnetic helicity can be produced.
Therefore, magnetic field with negative helicity is generated
simultaneously at larger scales.
This can also be understood as a result of mean-field dynamo theory,
where the sign of magnetic helicity at large scales agrees with the
sign of $\alpha_{\rm dyn}$ which, in turn, is a negative multiple
of the kinetic helicity \cite{Mof78,KR80}.

At later times, the magnetic helicity at small scales gets dissipated
resistivity, so that part of the magnetic helicity spectrum is
gradually lost until the entire magnetic helicity spectrum has
the same sign (negative) at all $k$.
The kinetic helicity has then also reversed sign, but it is
very small and sustained only by the current helicity.
After that time, the magnetic energy spectrum shows an inverse cascade
during which $\bra{\BB^2}\xiM^{1+\betaM}\approx\const$ with $\betaM\to0$;
see \Eq{Eq13}.

These new insights affect our understanding of all cases of decaying
turbulence with initial kinetic helicity in electrically conducting media,
such as plasma and liquid metal experiments, specifically the braked torus
experiment, neutron stars, galaxy clusters, inertial fusion confinement
plasmas, and the early universe.
Thus, we predict that experiments should approach an evolutionary track
in the $pq$ diagram close to the $\pM\approx1/2$ line for many tens of
thousands of turnover times if $\Rm$ is large enough.
Regarding applications to the early universe, the evolution in the
$\bra{\BB^2}$--$\xiM$ diagram (see Fig.~11 of Ref.~\cite{BKMRPTV17})
will be slightly steeper than for an initially fully helical
magnetic field.
This is because $\betaM$ is already close to zero.

We recall that kinetic and two-fluid effects have been neglected in the
present work.
While this should be appropriate for liquid metal experiments and the early
universe, it may not be accurate for plasma experiments and galaxy clusters.
Even in the dense neutron stars, Hall drift may play a role \citep{RG02}.
At present, not much is known about the importance of kinetic and
two-fluid effects for plasma decay in the presence of helicity,
so there is still a lot of room for basic studies in the field.

The new evolutionary phase of decaying magnetic fields with initial
kinetic helicity follows after an early phase of an exponential increase
of the magnetic energy by dynamo action.
It is here for the first time that such a process has been simulated.
To be able to achieve this, it was necessary to reach rather large values
of $\Rm$.
The subsequent decay phase with $\pM\approx1/2$ has so far only been
seen in the decaying phase of such a dynamo process.
During that time, the magnetic energy is already decaying, but the
system clearly captures signatures of the initial kinetic helicity in
the system, which then, at later times, disappears in favor of producing
first current helicity and later magnetic helicity.

The phenomenon of magnetic field amplification at intermediate times is
a new phenomenon specific to high magnetic Reynolds numbers, which are
only now becoming accessible to simulations.
At present, no detailed comparison with dynamo decay experiments is
possible yet, because no time-dependence of the magnetic field has
been obtained.
The best such experiment is that of two laser beams producing colliding
plasma jets directed toward each other, leading to magnetic field
generation that can be monitored through Faraday rotation measurements
\cite{Tzeferacos17b}.
The situation is complicated further by the fact that in the experiments
performed so far, the build-up phase of the turbulence constitutes a
significant fraction of the total time available.
One might therefore want to consider a model for the build-up of the
turbulence as well, which has not yet been attempted.

\begin{acknowledgments}
We thank Eric Blackman and two anonymous referees for useful comments.
AB acknowledges the University of Colorado's support through the
George Ellery Hale visiting faculty appointment.
TK acknowledges the High Energy and Cosmology Division and Associate
Membership Program at International Center for Theoretical Physics
(Trieste, Italy) for hospitality and partial support.
We also thank Nordita for hospitality during the programs
on Cosmological Magnetic Fields in 2015 (AB, TK, AGT, TV) and on
Chiral Magnetic Phenomena in 2018 (AB, TK, SM, ARP, TV).
Support through the NSF Astrophysics and Astronomy Grant (AAG) Program
(grants  AST1615940 \& AST1615100),
the Research Council of Norway (FRINATEK grant 231444),
the Swiss NSF SCOPES (grant IZ7370-152581), and the Georgian Shota Rustaveli
NSF (grant FR/264/6-350/14) are gratefully acknowledged.
TV is supported by the U.S.\ Department of Energy,
Office of High Energy Physics, under Award No.\ DE-SC0013605
at Arizona State University.
We acknowledge the allocation of computing resources
provided by the Swedish National Allocations Committee at the Center for
Parallel Computers at the Royal Institute of Technology in Stockholm.  This
work utilized the Janus supercomputer, which is supported by the National
Science Foundation (award No.\ CNS-0821794), the University of Colorado
Boulder, the University of Colorado Denver, and the National Center for
Atmospheric Research. The Janus supercomputer is operated by the University of
Colorado Boulder.
\end{acknowledgments}

\bibstyle{aps}
\bibliography{paper}

\end{document}